\documentclass[12pt]{article}
  \usepackage{amsfonts}
  \usepackage{amsmath}
\usepackage{amssymb}
\usepackage{amscd}
\usepackage[dvips]{graphicx}
\begin{document}
~~
\bigskip
\bigskip
\begin{center}
{\Large {\bf{{{Coherent states and quantum numbers for twist-deformed oscillator model }}}}}
\end{center}
\bigskip
\bigskip
\bigskip
\begin{center}
{{\large ${\rm {Marcin\;Daszkiewicz}}$}}\\
\bigskip
{ ${\rm{Institute\; of\; Theoretical\; Physics}}$}

{ ${\rm{ University\; of\; Wroclaw,\; pl.\; Maxa\; Borna\; 9,\;
50-206\; Wroclaw,\; Poland}}$}

{ ${\rm{ e-mail:\; marcin@ift.uni.wroc.pl}}$}
\end{center}
\begin{center}
{{\large ${\rm {Cezary\;J.\;Walczyk}}$}}\\
\bigskip
{ ${\rm{Department\; of\; Physics}}$}

{ ${\rm{ University\; of\; Bialystok,\; ul.\; Lipowa\;41,\; 15-424\; Bialystok,\; Poland}}$}

{ ${\rm{ e-mail:\; c.walczyk@alpha.uwb.edu.pl}}$}
\end{center}
\bigskip
\bigskip
\bigskip
\bigskip
\bigskip
\bigskip
\bigskip
\bigskip
\bigskip
\begin{abstract}
The coherent states for twist-deformed oscillator model provided in article \cite{daszwal} are constructed. Besides, it is demonstrated
that the energy spectrum of considered model is labeled by two quantum numbers - by so-called main and azimutal quantum numbers respectively.
\end{abstract}
\bigskip
\bigskip
\bigskip
\bigskip
\bigskip
\bigskip
\bigskip
\bigskip
\bigskip
 \eject

The suggestion to use noncommutative coordinates goes back to
Heisenberg and was firstly  formalized by Snyder in \cite{snyder}.
Recently, there were also found formal  arguments based mainly  on
Quantum Gravity \cite{2}, \cite{2a} and String Theory models
\cite{recent}, \cite{string1}, indicating that space-time at Planck
scale  should be noncommutative, i.e. it should  have a quantum
nature. Consequently, there appeared a lot of papers dealing with
noncommutative classical and quantum  mechanics (see e.g.
\cite{mech}, \cite{qm}) as well as with field theoretical models
(see e.g. \cite{prefield}, \cite{field}), in which  the quantum
space-time is employed.

In accordance with the Hopf-algebraic classification of all
deformations of relativistic \cite{clas1} and nonrelativistic
\cite{clas2} symmetries, one can distinguish three basic types
of space-time noncommutativity:\\
\\
{\bf 1)} The canonical (soft) deformation
\begin{equation}
[\;{ x}_{\mu},{ x}_{\nu}\;] = i\theta_{\mu\nu}\;, \label{noncomm}
\end{equation}
with constant and antisymmetric tensor $\theta_{\mu\nu}$. The
explicit form of corresponding Poincare Hopf algebra has been
provided in \cite{oeckl}, \cite{chi}, while  its nonrelativistic
limit  has been
proposed in \cite{daszkiewicz}. \\
\\
{\bf 2)} The Lie-algebraic  case 
\begin{equation}
[\;{ x}_{\mu},{ x}_{\nu}\;] = i\theta_{\mu\nu}^{\rho}x_{\rho}\;,
\label{noncomm1}
\end{equation}
with  particularly chosen constant coefficients
$\theta_{\mu\nu}^{\rho}$. Particular kind of such  space-time
modification has been obtained as representations of
$\kappa$-Poincare \cite{kappaP1}, \cite{kappaP2} and
$\kappa$-Galilei \cite{kappaG} Hopf algebras. Besides, the
Lie-algebraic twist deformations of relativistic and nonrelativistic
symmetries have been provided in \cite{lie1}, \cite{lie2} and
\cite{daszkiewicz}, respectively.
\\
\\
{\bf 3)} The quadratic deformation
\begin{equation}
[\;{ x}_{\mu},{ x}_{\nu}\;] =
i\theta_{\mu\nu}^{\rho\tau}x_{\rho}x_{\tau}\;, \label{noncomm2}
\end{equation}
with constant coefficients $\theta_{\mu\nu}^{\rho\tau}$. Its
Hopf-algebraic realization was proposed in \cite{qdef}, \cite{paolo}
and \cite{lie2}  in the case of relativistic symmetry, and in
\cite{daszkiewicz2} for its nonrelativistic counterpart.\\
\\
Besides, it has been demonstrated in \cite{nnh}, that in the case of
so-called N-enlarged Newton-Hooke Hopf algebras
$\,{\mathcal U}^{(N)}_0({ NH}_{\pm})$, the twist deformation
provides the new  space-time noncommutativity of the
form\footnote{$x_0 = ct$.},\footnote{ The discussed space-times have been  defined as the quantum
representation spaces, so-called Hopf modules (see e.g. \cite{oeckl}, \cite{chi}), for quantum N-enlarged
Newton-Hooke Hopf algebras.}
\begin{equation}
{ \bf 4)}\;\;\;\;\;\;\;\;\;[\;t,{ x}_{i}\;] = 0\;\;\;,\;\;\; [\;{ x}_{i},{ x}_{j}\;] = 
if_{\kappa \pm}\left({t}\right)\theta_{ij}(x)
\;, \label{nhspace}
\end{equation}
with time-dependent  functions
$$f_{\kappa +}\left({t}\right) = \kappa
f\left(\sinh\left(\frac{t}{\tau}\right),\cosh\left(\frac{t}{\tau}\right)\right)\;\;\;,\;\;\;
f_{\kappa -}\left({t}\right) = \kappa
f\left(\sin\left(\frac{t}{\tau}\right),\cos\left(\frac{t}{\tau}\right)\right)\;,$$
$\theta_{ij}(x) \sim \theta_{ij} = {\rm const}$ or
$\theta_{ij}(x) \sim \theta_{ij}^{k}x_k$ and  $\tau$ as well as $\kappa$ denoting  the cosmological constant and deformation parameter respectively.
It should be also noted that different relations  between all mentioned above quantum spaces ({\bf 1)}, { \bf 2)}, { \bf 3)}
and { \bf 4)}) have been summarized in article  \cite{nnh}.

Let us now turn to the quantum oscillator model defined on the twist-deformed phase space \cite{daszwal}\footnote{See type {\bf 4)} of quantum space-time.}
\begin{equation}
[\;t,{ {\bar x}}_{i}\;] = 0 \;\;\;,\;\;\;[\;{ {\bar x}}_{1},{ {\bar
x}}_{2}\;]  = if_{\kappa}(t)\;\;\;,\;\;\;
[\;{ {\bar x}}_{i},{\bar p}_j\;] = i\hbar\delta_{ij}\;\;\;,\;\;\;[\;{
{\bar p}}_{i},{ {\bar p}}_{j}\;] = 0\;. \label{phase}
\end{equation}
It's dynamic is given by the following Hamiltonian function with constant mass $m$ and frequency $\omega$
\begin{equation}
{\bar{H}}({\bar p},{\bar x})=\frac{1}{2m}\left({\bar p}_{1}^2 +
{\bar p}_{2}^2\right) +
\frac{1}{2}{m\omega^2}\left({\bar x}_{1}^2 + {\bar x}_{2}^2\right)\;, \label{hamosc}
\end{equation}
In order to analyze the above system we represent the
noncommutative variables $({\bar x}_i, {\bar p}_i)$ on classical
phase space $({ x}_i, { p}_i)$ as follows (see e.g. \cite{lumom},
\cite{kijanka})
\begin{equation}
{\bar x}_{1} = \hat{x}_{1} - \frac{f_{\kappa}(t)}{2\hbar}
\hat{p}_2\;\;\;,\;\;\;{\bar x}_{2} = \hat{x}_{2} +\frac{f_{\kappa}(t)}{2\hbar}
\hat{p}_1\;, \label{rep}
\end{equation}
where
\begin{equation}
[\;\hat{x}_i,\hat{x}_j\;] =0 =[\;\hat{p}_i,\hat{p}_j\;]
\;\;\;\;,\;\;\;\; [\;\hat{x}_i,\hat{p}_j\;] =i\hbar\delta_{ij}
\;.\label{ccr}
\end{equation}
Then, the  Hamiltonian (\ref{hamosc}) takes the form\footnote{It should be noted that for $f_{\kappa}(t) = \theta$ we get the canonically deformed oscillator model provided in \cite{kijanka}.}
\begin{eqnarray}
H_f(t) =\frac{\left({ \hat{p}}_{1}^2 + {\hat{p}}_{2}^2\right)}{2M_f(t)}
+\frac{1}{2}{M_f(t)\Omega_f^2(t)}\left({\hat{x}}_{1}^2 +
{\hat{x}}_{2}^2\right)-\frac{f_{\kappa}(t)}{2\hbar}m\omega^2\hat{L}\;, \label{genhamoscnew}
\end{eqnarray}
with symbol
\begin{equation}
\hat{L} = \hat{x}_1 \hat{p}_2 - \hat{x}_2 \hat{p}_1\label{momentum}
\end{equation}
 denoting angular momentum of particle. Besides, 
 the coefficients
 $M_f(t)$ and $\Omega_f (t)$ present in the  above formula  denote the
 time-dependent  functions given by
\begin{equation}
M_f(t)= \frac{m}{1 + \frac{m^2\omega^2f_{\kappa}^2(t)}{4\hbar^2}}
\;\;\;,\;\;\;\Omega_f (t)= \omega \sqrt{1 +
\frac{m^2\omega^2f_{\kappa}^2(t)}{4\hbar^2}}\;, \label{massfre}
\end{equation}
respectively, such that
\begin{equation}
M_f(t) \Omega_f^2 (t)= m \omega^2 = {\rm const.}\;. \label{prawo}
\end{equation}
Further, we introduce a set of time-dependent creation
$(a^{\dag}_{A}(t))$ and annihilation
$(a_{A}(t))$ operators 
\begin{eqnarray}
\hat{a}_{\pm}(t) &=& \frac{1}{2\sqrt{\hbar}}\left[\frac{(\hat{p}_1 \pm
i\hat{p}_2)}{\sqrt{M_f(t)\Omega_f (t)}} -i\sqrt{M_f(t)\Omega_f
(t)}(\hat{x}_1 \pm
i\hat{x}_2)\right]\;,\label{oscy1}
\end{eqnarray}
 satisfying the standard commutation relations
\begin{eqnarray}
[\;\hat{a}_{A},\hat{a}_{B}\;] =
0\;\;,\;\;[\;\hat{a}^{\dag}_{A},\hat{a}^{\dag}_{B}\;]
=0\;\;,\;\;[\;\hat{a}_{A},\hat{a}^{\dag}_{B}\;] =
\delta_{AB}\;\;\;;\;\;A,B = \pm\;.\label{ccr1}
\end{eqnarray}
Then, one
can easily check that in  terms of the  operators (\ref{oscy1}) the Hamiltonian function (\ref{genhamoscnew}) looks as
follows
\begin{equation}
{\hat{H}}_f(t)=\Omega_{+}(t) \left({\hat N}_+(t) + \frac{1}{2}\right)
+ \Omega_{-}(t) \left({\hat N}_-(t) + \frac{1}{2}\right)  \;, \label{hamquantosc}
\end{equation}
with
\begin{equation}
\Omega_{\pm}(t)=\Omega_f(t)\mp \frac{f_{\kappa}(t){m\omega^2
}}{2\hbar}\;,\label{ompm}
\end{equation}
and number operators in $\pm$ direction given by
\begin{equation}
{\hat N}_{\pm}(t)={\hat a}^{\dag}_{\pm}(t){\hat
a}_{\pm}(t)\;,\label{nn}
\end{equation}
respectively. Moreover, we see that
the energy eigenvectors can be  generated in a standard way as follows
\begin{eqnarray}
|n_+,n_-,t> =
\frac{1}{\sqrt{n_+!}}\frac{1}{\sqrt{n_-!}}\left({\hat
a}^{\dag}_{+}(t)\right)^{n_+} \left({\hat
a}^{\dag}_{-}(t)\right)^{n_-}|0>\;.\label{state}
\end{eqnarray}
while the corresponding (parameterized by $n_+$ and $n_-$)  eigenvalues are
\begin{equation}
E_{n_+,n_-}(t) = \Omega_{+}(t) \left(n_+ + \frac{1}{2}\right) +
\Omega_{-}(t) \left(n_- + \frac{1}{2}\right), \;\;\;n_+,n_- = 0, 1, 2, \ldots\;. \label{eigenvalues}
\end{equation}
Besides, using operator representation (\ref{oscy1}) one finds
\begin{equation}
(\Delta \hat{x}_i)^2_{|n_{+},n_{-},t>}(\Delta \hat{p}_i)^2_{|n_{+},n_{-},t>}= \frac{\hbar^2}{4}(1+n_{+}+n_{-})^2\;,\label{thhor}
\end{equation}
where symbol $(\Delta \hat{a})_{|\varphi>}$ denotes the uncertainty of observable $\hat{a}$ in quantum state $|\varphi>$.
The above result means that momentum-position uncertainty relations for eigenstates (\ref{state}) become saturated only for $n_+=n_-=0$, i.e. only for
vacuum vector $|0>$. Apart from that it is easy to see that the momentum operator (\ref{momentum}) can be written as follows
\begin{equation}
\hat{L}=\hbar\left(\hat{a}_{-}^\dagger(t) \hat{a}_{-}(t)-\hat{a}_{+}^\dagger(t) \hat{a}_{+}(t)\right)\;.\label{mmm}
\end{equation}
while it's action on quantum states (\ref{state}) is given by
\begin{equation}
\hat{L}|n_{+},n_{-},t>=\hbar(n_{-}-n_{+})|n_{+},n_{-},t>\;.\label{acttion}
\end{equation}
Consequently, the energy spectrum (\ref{eigenvalues}) can be written in terms of eigenvalues (\ref{acttion}) as follows
\begin{equation}
E_{n_+,n_-}(t)=\hbar \Omega_f(t)(n_++n_-+1)+\frac{f_{\kappa}(t) M_f(t)\Omega^2_f(t)}{2}(n_--n_+)\;.\label{spec}
\end{equation}

Let us now  solve two problems. First of them concerns the construction of so-called coherent states for considered model, i.e.
the quantum vectors which saturate the momentum-position Heisenberg uncertainty relations. The second problem applies to the proper interpretation of quantum
numbers $n = n_+ + n_-$ and $l = n_- -n_+$ labeling  the energy spectrum (\ref{spec}).

Hence, let us  consider the quantum states of the form
\begin{eqnarray}
|c_+,c_-,t> = \sum_{n_+,n_-}\frac{c_{+}^{n_+}{{\rm e}^{-\frac{1}{2}|c_+|^2}}}{\sqrt{n_+!}}
\frac{c_{-}^{n_-}{{\rm e}^{-\frac{1}{2}|c_-|^2}}}{\sqrt{n_-!}}
|n_+,n_-,t>\;,\label{coherent}
\end{eqnarray}
which play the role of  the eigenfunctions for annihilation operators (\ref{oscy1})
\begin{eqnarray}
{\hat a}_{\pm}(t)|c_+,c_-,t> = c_{\pm} |c_+,c_-,t>\;. \label{thor1}
\end{eqnarray}
 By direct calculation one may check that
\begin{eqnarray}
(\Delta p_i)_{|c_+,c_-,t>}^2 = \frac{\hbar M_f(t)\Omega_f(t)}{2}\;\;\;,\;\;\;  (\Delta
x_i)_{|c_+,c_-,t>}^2 = \frac{1}{2}\frac{\hbar}{M_f(t) \Omega_f(t)}\;,\quad i=1,2\;,
\end{eqnarray}
what leads to  the saturated momentum-position Heisenberg uncertainty relations
\begin{eqnarray}
(\Delta p_i)_{|c_+,c_-,t>}^2 (\Delta x_i)_{|c_+,c_-,t>}^2 = \frac{\hbar^2}{4}\;, \quad i=1, 2\;.\label{nieoznaczonosc}
\end{eqnarray}
Consequently, we see that the vectors (\ref{coherent}) are (in fact) nothing else than the coherent states for twist-deformed oscillator model,
satisfying
\begin{equation}
<\hat{H}_f>_{|c_{+},c_{-},t>}=E_{|0,0,t>}(t)+\frac{\Omega_f(t)}{\hbar}(\Delta L)^2_{|c_{+},c_{-},t>}+ \frac{M_f(t)\Omega_f^2(t) f_{\kappa}(t)}{2\hbar}<L>_{|c_{+},c_{-},t>}\;,\label{odyn}
\end{equation}
with
\begin{eqnarray}
<L>_{|c_{+},c_{-},t>} &=&\hbar(|c_{-}|^2-|c_{+}|^2)\;,\\
(\Delta L)^2_{|c_{+},c_{-},t>}&=&\hbar^2(|c_{-}|^2+|c_{+}|^2)\;.
\end{eqnarray}

In the case of second problem one should to solve the eigenvalue
 equation for Hamiltonian (\ref{genhamoscnew})
written in terms of  polar coordinates
\begin{equation}
\hat{H}_f(t)\psi(r,\varphi,t) = E(t)\psi(r,\varphi,t)\;,\label{polar}
\end{equation}
where
\begin{eqnarray}
\hat{H}_f(t)&=&-\frac{\hbar^2}{2M_f(t)}\left(\frac{\partial^2}{\partial r^2}+\frac{1}{r}\frac{\partial}{\partial r}-\frac{1}{\hbar^2}\frac{\hat{L}^2}{r^2}\right)\;+
\label{polar}\\
&~~&~~~~~~~~~~~~~~~~~~~~~+\;\frac{M_f(t)\Omega_f^2(t)}{2}r^2-\frac{f_{\kappa}(t) M_f(t) \Omega_f^2(t)}{2\hbar}\hat{L}\;,\nonumber
\end{eqnarray}
and
\begin{equation}
\hat{L}=-i\hbar \frac{\partial}{\partial \varphi}\;\;\;,\;\;\; [\hat{H},\hat{L}]=0\;.\label{l}
\end{equation}
To this aim, it is convenient to take the corresponding eigenfunctions   in the form
\begin{equation}
\psi(r,\varphi,t)=\phi(\varphi)R(r,t)\;,\label{solution}
\end{equation}
with it's azimutal  part $\phi(\varphi)$ satisfying
\begin{equation}
\hat{L}\phi_l(\varphi)=\hbar l\phi(\varphi)\;\;\;,\;\;\; \phi_l(\varphi)=\frac{1}{\sqrt{2\pi}}\mathrm{e}^{il\varphi}\;,\quad l=0,\pm 1,\pm 2,
\dots\;.\label{katowa}
\end{equation}
Then, the proper equation for radial function $R(r,t)$ looks as follows
\begin{eqnarray}
&~~&\left(-\frac{\partial^2}{\partial \rho^2}-\frac{1}{\rho}\frac{\partial}{\partial \rho}+\frac{l^2}{\rho^2}+\frac{\rho^2}{4}-\mathcal{E}_l(t)\right)R_l(\rho(t))=0\;,\label{radial} \\
&~~&\mathcal{E}_l(t)=\frac{E(t)-\frac{f_{\kappa}(t) M_f(t) \Omega_f^2(t)}{2}l}{\hbar \Omega_f(t)}\;,\nonumber
\end{eqnarray}
where  $\rho(t)=r\sqrt{2M_f(t)\Omega_f(t)/ \hbar}$ plays the role of dimensionless variable. It's physical solution can be written as
\begin{equation}
R_l^{(n)}(\rho(t))=w_l^{(n)}(\rho(t))\mathrm{e}^{-\rho^2(t)/4}\;,\label{standard}
\end{equation}
with $w_l^{(n)}(\rho(t))$ denoting the polynomial of degree $n$. Then, the equation (\ref{radial}) reduces to the following one
\begin{eqnarray}
&&-\frac{\partial^2 w_l^{(n)}(\rho(t))}{\partial\rho^2}+\frac{\rho^2-1}{\rho}\frac{\partial w_l^{(n)}(\rho(t))}{\partial\rho}\;+
\label{redukcja}\\
&~~&~~~~~~~~~~~~~~~~~~~~~~~~~~+\;\frac{l^2}{\rho^2}w_l^{(n)}(\rho(t))-(\mathcal{E}_l(t)-1) w_l^{(n)}(\rho(t))=0\;,\nonumber
\end{eqnarray}
for which the solution (this time) is given by\footnote{The symbol $a_l^{(n)}$ denotes the normalization factor.}
\begin{equation}
w_l^{(n)}(\rho(t))=a_l^{(n)}\left(1+\sum_{k=1}^{(n-|l|)/2}\left[\prod_{s=1}^k\frac{n+2-(2s+|l|)}{l^2-(2s+|l|)^2}\right]\rho^{2k}(t)\right)\rho^{|l|}(t)\;,
\label{roz}
\end{equation}
only when
\begin{equation}
\mathcal{E}_l(t)\rightarrow \mathcal{E}_l^{(n)}(t)=n+1,\quad l\in\{-n,-n+2,\dots,n-2,n\},\quad n=0, 1, 2, 3, \dots\;,\label{warunki}
\end{equation}
or (equivalently)
\begin{equation}
E_l^{(n)}(t)=\hbar \Omega_f(t)(n+1)+\frac{f_{\kappa}(t) M_f(t)\Omega_f^2(t)}{2}l\;.\label{spec2}
\end{equation}
Consequently, after substitution $n=n_{+}+n_{-}$ and $l=n_{-}-n_{+}$ into eigenvalues (\ref{spec2}) we get (in fact) the energy spectrum (\ref{spec})
labeled by $n_+$ and
$n_-$ parameters. For this reason as well as  due to the formulas (\ref{katowa}), (\ref{standard}) and (\ref{spec2})  the quantities $n$ and $l$ may be called  the ''main'' and ''azimutal'' quantum numbers respectively.


\section*{Acknowledgments}
The authors would like to thank J. Lukierski
for valuable discussions.\\
This paper has been financially supported by  Polish
NCN grant No 2011/01/B/ST2/03354.

\end{document}